\documentstyle[12pt]{article}
\begin{document}
\thispagestyle{empty}
\begin{raggedleft}
\end{raggedleft}
$\phantom{x}$\vskip 0.618cm\par
\begin{center}{\Huge Destructive Interference of Dualities}
\vspace{2cm}
$\phantom{X}$\\
{\Large Clovis Wotzasek
}\\[3ex]
{\em Instituto de F\'\i sica\\
Universidade Federal do Rio de Janeiro\\
21945, Rio de Janeiro, Brazil\\}
and\\
{\em Department of Physics and Astronomy\\
University of Rochester\\
Rochester, NY 14627, USA}
\end{center}\par
\vspace{2cm} 
\abstract

\noindent The soldering mechanism
has been shown to represent the quantum interference
effect between self and anti-self dual aspects of a given symmetry.
This mechanism was used to show that the massive mode of the 2D
Schwinger model results from the constructive interference
between the right and the left massless modes of chiral
Schwinger models.  Similarly, the topologically massive modes
resulting from the bosonization of 3D massive Thirring models of
opposite mass signatures, are fused into the two massive modes of the
3D Proca model, thanks to the interference of dualities characteristic
of the soldering mechanism.  In this work, we show that the
field theoretical analog of destructive
quantum mechanical interference may also be represented by the soldering
mechanism.  This phenomenon is illustrated by the fusion of two
(diffeomorphism) invariant
self-dual scalars described by right and left chiral-WZW
actions, producing a Hull non-mover field. After fusion,
right and left moving modes disappear from the spectrum,
displaying the claimed (destructive) interference of dualities.          
          
\vspace{2cm}

\newpage
The investigation in chiral bosonization has started many years ago
in the seminal work of Siegel\cite{WS}.  Later on Gates and Siegel
showed how to construct general interacting actions for chiral bosons,
including the supersymmetric and the non-abelian cases\cite{GS}.
They used this construction to obtain the righton-lefton interaction
by carrying out the path integral quantization in a generalized Thirring
model.
Alternatively, Stone\cite{MS} has shown that the method of 
coadjoint orbit when applied to a representation of a group
associated with a single affine Kac-Moody algebra, provides an action for
the chiral WZW model\cite{EW}, a non-abelian
generalization of the Floreanini and Jackiw model\cite{florjack}. 
This method gives an useful
bosonization scheme for Weyl fermions, since a level one representation
of LU(N) has an interpretation as the Hilbert space for a free chiral
fermion\cite{PS}.  The drawback is that only Weyl
fermions can be dealt with in this way, since a 2D conformally invariant
QFT has separate right and left current algebras.  
In order to overcome this difficulty, Stone\cite{MS} introduced the idea of
soldering the two chiral scalars by introducing a non-dynamical gauge field to
remove the degree of freedom that obstructs the vector gauge invariance. 
This is connected to the observation that one needs more than the 
direct sum of two fermionic
representations of the Kac-Moody algebra to describe a Dirac fermion.
Stated differently, the equality for the weights in the two representations
is physically connected with the need to abandon one of the two separate
chiral symmetries, and accept that only vector gauge symmetry should be
maintained.  This is the main motivation for the introduction of the
soldering field which pervades for the fusion of dualities
in all space-time dimensions.  Moreover, being just an auxiliary field,
it may posteriorly
be eliminated in favor of the physically relevant quantities.
This restriction will force the two independent chiral representations
to belong to the same multiplet, {\it effectively} soldering them together.

On the other hand, the role of duality as a qualitative tool in the investigation
of physical systems is being gradually disclosed at different
context and dimensions\cite{AG}. 
In two space-time dimensions in particular,
we face the intriguing situation where chirality also plays the role
of duality. 
This enables the investigation of the former to be performed
using the techniques developed for the latter. 
Recently, this author and collaborators\cite{BW,BW2,BW3} extended
the techniques of fusion
or soldering, introduced by Stone\cite{MS}, to investigate some
new aspects of dualities at different space-time dimensions, and studied 
the physical consequences of their
combination by the soldering process.
In particular, we have shown\cite{BW} that the constructive interference
between the left
and the right moving massless modes of two chiral Schwinger
models\cite{JR} of 
opposite chiralities,
are soldered into the gauge invariant massive mode of the vector
Schwinger model. In fact, by equipping the soldering technique with
gauge and Bose symmetry\cite{NR} automatically
select the massless sector of the chiral models displaying
the Jackiw-Rajaraman parameter that reflects
the bosonization ambiguity by $a=1$.
In the 3D case, the soldering mechanism was used to show the
result of fusing together two topologically massive modes
generated by the bosonization of two
massive Thirring models with opposite mass signatures in the long
wave-length limit.  The bosonized modes, which are described by
self and anti-self dual Chern-Simons models\cite{PRN,DJ}, were
then soldered into
the two massive modes of the 3D Proca model\cite{BW2}.
In the 4D case, the soldering mechanism produced an explicitly
dual and covariant action as the result of the interference between 
two Schwarz-Sen\cite{SS} actions displaying opposite aspects of
the electromagnetic duality\cite{BW3}.  It is our intention in this work 
to study the physical consequences of combining
actions possessing truncate diffeomorphism invariance and
opposite chiralities using the fusion of dualities technique.

It should be mentioned that such procedure has a typical quantum
mechanical nature, with no classical parallel.  It is completely
meaningless to perform the sum of two classical actions, that although
describing opposite aspects of some (duality) symmetry, would depend
on the same field.  On the other hand, the direct sum of duality symmetric
actions depending on different fields would not give anything new.  It is
the soldering process that leads to a new and non trivial
result.
In 2D, this result has been interpreted\cite{BW} as the consequence of the
constructive
interference of chiralities, by coupling the chiral
scalar fields to
a dynamical gauge field. The resulting effective gauge theory, obtained after
the elimination of the soldered scalar field through equations of motion, shows
the presence of a mass term that is typical of the right-left quantum
interference\cite{RJ}.

In this work we show that it is also possible to obtain the field
theoretical analogue of the
``quantum destructive interference" phenomenon, by coupling the chiral
scalars to appropriately truncated metric fields, known as
chiral WZW models, or non abelian Siegel models.  By soldering the 
two (Siegel) invariant representations
of the chiral WZW model\cite{WS} of opposite chiralities, the effective
action that results from this process is
shown to be invariant under the full diffeomorphism group, which is
not a mere sum of two Siegel symmetries. In fact, this effective action
does not contain either right or left movers, but can be
identified with the non-abelian generalization of the
bosonic non-mover action proposed by Hull\cite{CH},
thanks to the richer symmetry structure
induced over it by the soldering mechanism.

To begin with, let us review some facts about the
non-abelian Siegel model\cite{frishson}.  The action for
a left mover chiral scalar
is given as\footnote{Our notation is as follows: $x^{\pm}={1\over 2}(t\pm x)$
are the light-cone variables and $\tilde g = g^{-1}$ denotes the inverse
matrix.},

\begin{equation}
\label{leftzero}
S_0^{(+)}(g)=\int\;d^2x\; tr\left(\partial_+g\:
\partial_-\tilde g+\lambda_{++}\partial_-g\:\partial_-\tilde g\right)
+\Gamma_{WZ}(g)\, ,
\end{equation}

\noindent where $g\in G$ is a matrix-valued field taking values on some compact
semi-simple Lie group $G$, with an algebra ${\hat G}$. 
The term $\Gamma_{WZ}(g)$ is the topological Wess-Zumino functional,
as defined in Ref.\cite{polywieg}.  It is invariant under a chiral
diffeomorphism known as Siegel transformation where,

\begin{equation}
\label{siegel}
\delta\lambda_{++} = -\partial_+\epsilon^- + \lambda_{++}
\partial_-\epsilon^- +\epsilon_-\partial_-\lambda_{++}\, ,
\end{equation}

\noindent and $g$ transforming as a scalar.  This action can be
seen as the
WZW action, immersed in a gravitational background, with an appropriately
truncated metric tensor,

\begin{equation}
\label{leftzerograv}
S_0^{(+)}(g)={1\over 2}\int d^2 x \sqrt{-\eta_+}\; 
\eta_+^{\mu\nu}\:tr\left(\partial_\mu g\:
\partial_\nu \tilde g\right) +\Gamma_{WZ}(g)\, ,
\end{equation}

\noindent with $\eta^+ = det(\eta^+_{\mu\nu})$ and

\begin{equation}
\label{metric+}
{1\over 2} \sqrt{-\eta_+}\: \eta_+^{\mu\nu}=\left(
\begin{array}{cc}
0 & {{1\over 2}}\\
{{1\over 2}} & {\lambda_{++}}
\end{array}
\right)\, .
\end{equation}

Next, let us compute the Noether currents for the axial, vectorial and the
right and left chiral transformations.  The variation of the Siegel-WZW
action (\ref{leftzero}) or (\ref{leftzerograv}) gives,

\begin{equation}
\label{variation}
\delta S_0^{(+)}(g)=\left\{
\begin{array}{lll}
\int d^2x\:tr\Big\{\delta g \tilde g\: 
2\Big[\partial_+\left(\partial_- g\tilde g\right)
\:+\:\partial_-\left(\lambda_{++}\partial_- g\tilde g\right)\Big]\Big\}\\
\mbox{}\\
\int d^2x\:tr\Big\{\tilde g\delta g \: 
2\Big[\partial_-\left(\tilde g\partial_+ g\right)
\:+\:\partial_-\left(\lambda_{++}\tilde g\partial_- g\right)\Big]\Big\}\, .
\end{array}\right.
\end{equation}

\noindent  From (\ref{variation}) and the axial transformation
($g\rightarrow kgk $) we obtain,

\begin{eqnarray}
\label{axialcurr}
J_A^+&=& 2g\partial_-\tilde g\nonumber\\
J_A^-&=& -2\Big[\tilde g\partial_+ g +\lambda_{++}
(\tilde g\partial_- g +\partial_-g \;\tilde g)\Big]\, ,
\end{eqnarray}

\noindent where $k \in K $ take their values in some
subgroup $K\subset G$.  From the transformation
($ g \rightarrow \tilde k g k$) we obtain the vector current,

\begin{eqnarray}
\label{vectcurr}
J_V^+&=& 2g\partial_-\tilde g\nonumber\\
J_V^-&=& 2\Big[\tilde g\partial_+ g +\lambda_{++}(\tilde g\partial_- g -
\partial_-g \;\tilde g)\Big]\, .
\end{eqnarray}

\noindent Incidentally, it should be observed that the axial and the vectorial
currents (\ref{axialcurr}) and (\ref{vectcurr}) are dual to each other
only if the following extended definition is adopted,

\begin{equation}
\mbox{}^*T^\mu=\sqrt{-\eta_+} \; \eta_+^{\mu\nu}
\epsilon_{\mu\lambda}T^\lambda\, ,
\end{equation}

\noindent and use of the following relations is made,
\begin{eqnarray}
J_+ &=& J^- -2 \lambda_{++}J^+\nonumber\\
J_- &=& J^+\, ,
\end{eqnarray}

\noindent which is valid for all currents.
Similarly, the chiral currents can be obtained  from the left
($g\rightarrow gk$) and
right ($g\rightarrow \tilde k g$) transformation. The result is ,

\begin{eqnarray}
J_L^{(+)} &=& 0\nonumber\\
J_L^{(-)} &=& 2\left(\tilde g \partial g +\lambda_{++} 
\tilde g\partial_- g\right)\, ,
\end{eqnarray}

\noindent and

\begin{eqnarray}
\label{direita}
J_R^{(+)} &=& -2 g\partial_-\tilde g\nonumber\\
J_R^{(-)} &=& -2\lambda_{++} g \partial_-\tilde g\, .
\end{eqnarray}

\noindent  It is crucial to notice that out of the two affine invariances
of the original WZW model, only one is left over due to the chiral constraint
$\partial_- g\approx 0$.  Indeed, the affine invariance is only present
in the left sector since $J_L^{(+)}=0$ and $\partial_-J_L^{(-)}=0$,
which implies $J_L^{(-)}=J_L^{(-)}(x^+)$,
while $J_R^{(-)}\neq 0$ and $J_R^{(+)}\neq J_R^{(+)}(x^-)$.

Next we work out the details for the right chirality action,

\begin{eqnarray}
\label{rightzero}
S_0^{(-)}(h)&=&\int \;d^2x\; tr\left(\partial_+ h
\partial_-\tilde h +\lambda_{--}\partial_+ h
\partial_+\tilde h\right) - \Gamma_{WZ}(h)\nonumber\\
&=& {1\over 2} \int d^2x\; \sqrt{-\eta_-} \;\eta_-^{\mu\nu}
\:tr\left(\partial_\mu h \partial_\nu 
\tilde h\right)- \Gamma_{WZ}(h)\, ,
\end{eqnarray}

\noindent  where

\begin{equation}
{1\over 2}\sqrt{-\eta_-}\:\eta_-^{\mu\nu}=\left(
\begin{array}{cc}
{\lambda_{--}} & {{1\over 2}}\\
{{1\over 2}} & 0
\end{array}
\right).
\end{equation}

\noindent  Notice that $S_0^{(+)}(g)$ and $S_0^{(-)}(h)$
are chosen at opposite critical points, otherwise they will
not carry different chiralities, a crucial condition
for the soldering to be performed.
The set of axial, vector and chiral Noether currents is similarly
obtained,

\begin{eqnarray}
J_A^{(+)}(h)&=& 2\Big[\tilde h\partial_- h +\lambda_{--}
\left(\tilde h\partial_+ h +\partial_+ h \tilde h\right)\Big]\nonumber\\
J_A^{(-)}(h)&=&2\:\partial_+ g\tilde g\\
J_V^{(+)}(h)&=&2\Big[\tilde h\partial_- h +\lambda_{--}
\left(\tilde h\partial_+ h -\partial_+ h\:\tilde h\right)\Big]\nonumber\\
J_V^{(-)}(h)&=&-2\:\partial_+ h\:\tilde h\\
J_L^{(+)}(h)&=& 2\:\partial_+ h\:\tilde h\nonumber\\
J_L^{(-)}(h)&=&2\:\lambda_{--}\:\partial_+ h\:\tilde h\\
J_R^{(+)}(h)&=&2\left(\tilde h\partial_- h +\lambda_{--}
\:\tilde h\partial_+ h\right)\nonumber\\
J_R^{(-)}(h)&=&0\, ,
\end{eqnarray}

\noindent with the corresponding interpretation analogous to
that following Eq.(\ref{direita}).  Since  the actions (\ref{leftzero})
and (\ref{rightzero}) do 
correspond to opposite aspects of a symmetry (chirality), the stage is set
for the soldering.

Next, let us discuss the gauging procedure to be adopted in the soldering
of the right and the left chiral WZW actions just reviewed.  
The basic idea of the soldering procedure is to lift a global Noether symmetry
present at each individual chiral component into a local symmetry for the
composite system that, consequently, defines the soldered action.
It is of vital importance to notice that
the coupling with the (auxiliary) soldering gauge field is only consistent
if use is made of the correspondent chiral current.
Otherwise the equations of motion, after the gauging, will result being incompatible with the covariant chiral constraint, by the presence of an anomaly.  Anomalies can
certainly, be accommodate into the theory, but not at the expense of violating
the consistence between equations of motion and gauge constraints.  Here we
shall adopt an iterative Noether procedure to lift the global
(left) chiral symmetry of (\ref{leftzero}),

\begin{eqnarray}
\label{transf.1}
g&\rightarrow & gk\nonumber\\
\lambda_{++}&\rightarrow & \lambda_{++}\nonumber\\
A_-&\rightarrow & \tilde k A_-k + \tilde k\partial_- k
\end{eqnarray}

\noindent  into a local one.  To compensate for the non-invariance
of $S_0^{(+)}$, we introduce the coupling term,

\begin{equation}
S_0^{(+)}\rightarrow S_1^{(+)} =S_0^{(+)} +A_- J_L^-(g)\, ,
\end{equation}

\noindent  along with the soldering gauge field $A_- $, taking values
in the subalgebra ${\hat K}$ of $K$, whose transformation properties
are being defined in (\ref{transf.1}).  Using such transformations,
it is a simple algebra to find that,

\begin{equation}
\delta\left(S_1^{(+)}-\lambda_{++} A_-^2\right)=2\;\partial_+\omega\; A_-\, ,
\end{equation}

\noindent with $\omega \in {\hat K}$ being an infinitesimal element of
the algebra.
One can see that ,
\begin{equation}
S_2^{(+)}=S_1^{(+)} -\lambda_{++}\;A_-^2
\end{equation}
cannot be made gauge invariant by additional Noether counter-terms, but
it has the virtue of being independent of the transformation
properties of $g$ while depending
only on the elements of the gauge algebra $\hat K$.  
Similarly, for the right chirality we find,

\begin{equation}
\delta S_2^{(-)}=-\;2\;A_+\;\partial_-\omega
\end{equation}

\noindent for

\begin{equation}
S_2^{(-)}(h)=S_0^{(-)}(h)-A_+J_R^{+}(h)-\;\lambda_{--}\;A_+^2
\end{equation}

\noindent  when the basic fields transform as,

\begin{eqnarray}
\label{transf.2}
h&\rightarrow &hk\nonumber\\
A_+&\rightarrow & kA_+\tilde k + k\partial_+\;\tilde k\nonumber\\
\lambda_{--} &\rightarrow & \lambda_{--}\, .
\end{eqnarray}

\noindent  It is important to observe that the action for the right
sector depends functionally on a different field, namely $h \in H$. 
Although the gauged actions for each chirality could not
be made gauge invariant separately,  with the inclusion of a contact
term, the combined action,

\begin{equation}
\label{eff}
S_{eff}=S_2^{(+)} + S_2^{(-)} + 2 A_+ \; A_-\, ,
\end{equation}

\noindent is invariant under the set of transformations
(\ref{transf.1}) and (\ref{transf.2}) simultaneously.

Following Ref.\cite{MS}, we eliminate the (non dynamical) gauge
field $ A_\mu $.  From the equations of motion one gets,

\begin{equation}
{\cal J} = 2 {\bf M} {\cal A}\, ,
\end{equation} 

\noindent where we have introduced the following matricial notation,

\begin{equation}
{\cal J} =\left(
\begin{array}{c}
J_L^-(g)\\
J_R^+(h)
\end{array}
\right)
\end{equation}

\begin{equation}
{\cal A} =\left(
\begin{array}{c}
{A_+}\\
{A_-}
\end{array}
\right)
\end{equation}

\noindent and

\begin{equation}
{\bf M} =\left(
\begin{array}{cc}
1 & {\lambda_{++}}\\
{\lambda_{--}} & 1
\end{array}
\right)\, .
\end{equation}

\noindent  Bringing these results into the effective action (\ref{eff}) gives,

\begin{eqnarray}
S_{eff} &=& S_0^{(+)}(g) + S_0^{(-)}(h) +\nonumber\\
&+& \int d^2x {1\over{1-\lambda^2}} 
tr\left\{ 2\left[\tilde g\partial_+ g \;\tilde h\partial_- h + 
\lambda^2 \;\tilde g\partial_- g\;\tilde h\partial_+ h 
+\right.\right.\nonumber\\
&+& \left. \lambda_{++}\;\tilde g\partial_- g \;\tilde h\partial_- h 
+\lambda_{--}\;\tilde g\partial_+ g\;\tilde h\partial_- h \right] +\nonumber\\
&+& \lambda_{--}\;\left(\partial_+ g\;\partial_+\tilde g + 2\lambda_{++}
\;\partial_+ g\;\partial_-\tilde g +\lambda_{++}^2\;\partial_- g\;
\partial_-\tilde g\right) +\nonumber\\
&+& \left. \lambda_{++}\;\left(\partial_- h\;\partial_-\tilde h + 
2 \lambda_{--}\;\partial_+ h\;\partial_+\tilde h 
+\lambda_{--}^2\;\partial_+ h\;\partial_+\tilde h\right)\right\}\, .
\end{eqnarray}

\noindent where $\lambda^2=\lambda_{++}\lambda_{--}$. 
It is now a simple algebra to show that this effective action
does not depend on the fields $g$ and $h$ individually, but
only on a gauge invariant combination of them, defined below
(\ref{sss}). 
This effective action corresponds to that of a (non-chiral) WZW model
coupled minimally to an effective metric built out of the Lagrange multipliers
of the original Siegel actions,

\begin{equation}
\label{ss}
{1\over 2}\sqrt{-\eta}\; \eta^{\mu\nu}= {1\over{1-\lambda^2}}\;\left(
\begin{array}{cc}
{\lambda_{--}} & {{{1+\lambda^2}\over 2}}\\
{{{1+\lambda^2}\over 2}} & {\lambda_{++}}
\end{array}
\right)
\end{equation}

\noindent and a new (effective) field,

\begin{equation}
\label{sss}
{\cal G}= g\tilde h
\end{equation}

\noindent and reads,

\begin{equation}
\label{s}
S={1\over 2}\int \:d^2x\:\sqrt{-\eta}\:\eta^{\mu\nu}\:tr\left(\partial_\mu 
{\cal G}\;\partial_\nu\tilde{\cal G}\right) +\Gamma_{WZ}({\cal G})\, .
\end{equation}

\noindent Here we have used the well known property of the Wess-Zumino
functional $\Gamma_{WZ}(h)\:=\:-\;\Gamma_{WZ}(\tilde h)$, and the
Polyakov-Weigman identity\cite{polywieg}.

It is interesting to notice that the original chiral transformations
(\ref{transf.1})
and (\ref{transf.2}) are now hidden, since the effective action is
composed of only the
gauge invariant objects (\ref{ss}) and (\ref{sss}).  To unravel the physical
contents of the effective soldered action (\ref{s}), it is important
to study the new set of symmetries
of the composite theory.  We first observe that under 
diffeomorphism the metric transform as a symmetric tensorial density,
                           
\begin{eqnarray}
\label{138}
\delta\lambda_{++}&=&-\partial_+\epsilon^-+\lambda^2_{++}\partial_-\epsilon^+
+(\partial_+\epsilon^+-\partial_-\epsilon^-+\epsilon^+\partial_+
+\epsilon^-\partial_-)
\lambda_{++}\nonumber\\
\delta\lambda_{--}&=&-\partial_-\epsilon^++\lambda^2_{--}\partial_+\epsilon^-
+(\partial_-\epsilon^--\partial_+\epsilon^++\epsilon^+\partial_+
+\epsilon^-\partial_-)
\lambda_{--}
\end{eqnarray}
while the ${\cal G}$ transforms as a scalar.
It is important to observe that if we
restrict the diffeomorphism to just one sector,
say by requiring $\epsilon^+=0$, 
we reproduce the original Siegel
symmetry for the sector described by the pair ${\cal G}\,,\lambda_{++}$
in the same way as it appears in the original chiral theory (\ref{leftzero}).
However,
under this restriction,
$\lambda_{--}$ transforms in a non-trivial
way as,

\begin{equation}
\label{c1}
\delta\lambda_{--}=\lambda^2_{--}\partial_+\epsilon^-+
\left(\partial_-\epsilon^-+\epsilon^-\partial_-\right)\lambda_{--}\,.
\end{equation}
The original Siegel symmetry, therefore, is 
not a subgroup of the diffeomorphism group but 
it is only recovered if we also make a 
further truncation, by imposing  that $\lambda_{--}=0$. 
The existence of the residual symmetry (\ref {c1}) seems to be related
to a duality symmetry satisfied by the effective action (\ref{s}) when the
metric is parametrized as in (\ref{ss}).
Under the discrete transformation,

\begin{equation}
\label{c2}
\lambda_{\pm\pm}\rightarrow{1\over\lambda_{\mp\mp}}  \,,
\end{equation}

\noindent the residual transformation (\ref{c1}) swaps to
(\ref{siegel}) while that becomes the residual symmetry for
the opposite chiral sector.  Indeed we see that the
classical equations of motion remain invariant
under (\ref{c2}) while the effective action changes its signature,
very much like in the original electromagnetic duality transformation.
This is obviously
related to the interchange symmetry between the right and the
left moving sectors
of the theory, and seems to be of general validity\cite{BW3}. 
Also notice that the gauged Lagrangian in one sector, either $S_2^+$ or
$S_2^-$,
cannot be written in a diffeomorphism invariant manner. Therefore,
gauging in one of the sectors breaks Siegel invariance. However,
let us note that if we integrate out either the $A_-$ or the $A_+$
field, the Siegel theory changes
chirality with the identification provided by (\ref{c2}), that is again
related to the discrete duality symmetry.

Now comes the crucial observation.  By solving the equations of motion
and setting the $\lambda_\pm$ to zero by invoking the diffeomorphism invariance
discussed above, it is simple to see that the composite field (\ref{sss})
of the effective
action (\ref{s})
describes a non-mover field, as first proposed by Hull\cite{CH}. 
The right and the left moving modes have therefore
disappeared from the spectrum.
The soldering procedure has clearly produced a destructive interference between
left and right movers of the original chiral components.  Moreover,
it can be easily seen that the
the coupling of chiral scalars to a dynamical gauge field
before soldering, as done in \cite{BW} (see appendix),
will decouple the gauge sector
from the effective soldered action.  This seems to be a
natural result since a non-mover field
cannot couple to either right or left components of
the vector gauge field.  This is a
distinctive result produced by the presence of the full group of diffeomorphism
resulting from the soldering process, that constrains the matter
scalar field into the non-moving sector, quite in opposition to the 
constructive interference result that comes from soldering the noninvariant
models.\vspace{0.3cm}\\

\noindent ACKNOWLEDGEMENTS.  The author would like to thank the members
of the Department of Physics and Astronomy of University of Rochester for the
hospitality during the visit that was made possible by the bilateral agreement
CNPq-NSF.  The author is partially supported by CNPq, FINEP and FUJB , Brazil.


\begin{thebibliography}{30}
\bibitem{WS}W. Siegel, Nucl.Phys. B238 (1984) 307.
\bibitem{GS} S.J. Gates Jr. and W. Siegel, Phys. Lett. B206 (1988) 631.
\bibitem{MS}M. Stone, Phys. Rev. Lett. 63 (1989) 731;
Nucl. Phys. B327 (1989) 399,
and Illinois Report, ILL-TH-89-23.
\bibitem{EW}E. Witten, Comm.Math.Phys. 92 (1984) 455.
\bibitem{florjack}R. Floreanini and R. Jackiw, Phys.Rev.Lett. 59 (1987) 1873.
\bibitem{PS}A. Pressley and G. Segal, "Loop Groups", Oxford 
University Press, Oxford, 1986.
\bibitem{AG}L. Alvarez-Gaume and S.F. Hassan, Fortsch. Phys. 45 (1997) 159;
Also see, L. Alvarez-Gaume and F. Zamora, Duality in Quantum Field theory
(and String Theory), hep-th/9709180.
\bibitem{BW} E.M.C. de Abreu, R. Banerjee and C. Wotzasek, Nucl. Phys. B509
(1998) 519.
\bibitem{BW2}R. Banerjee and C. Wotzasek, Bosonization and Soldering of
Dual Symmetries in Two and Three Dimensions, hep-th/9709105.
\bibitem{BW3}R. Banerjee and C. Wotzasek, Duality Symmetry and Soldering at
Different Dimensions, hep-th/9710060.
\bibitem{JR}R. Jackiw and R. Rajaraman, Phys. Rev. Lett. 54 (1985) 1219.
\bibitem{NR}N. Banerjee and R. Banerjee, Nucl. Phys. B445 (1995) 516.
\bibitem{PRN} P.K. Townsend, K. Pilch and P. van Nieuwenhuizen,
Phys. Lett. B136 (1984) 452.
\bibitem{DJ}S. Deser and R. Jackiw, Phys. Lett. B139 (1984) 371.
\bibitem{SS} J.Schwarz and A. Sen, Nucl. Phys. B411 (1994) 35.
\bibitem{RJ}R. Jackiw, Diverse Topics in Theoretical and
Mathematical Physics, World Scientific, Singapore, 1995.
\bibitem{frishson}Y. Frishman and J. Sonnenschein, Nucl.Phys. B301 (1988) 346.
\bibitem{CH}C. Hull, Phys.Lett.B206(1988)234.
\bibitem{polywieg}A. Polyakov and P.B. Weigman, 
Phys.Lett B131 (1983) 121; Phys.Lett. B141 (1984) 223.
\end{thebibliography}
\end{document}